\documentclass[12pt]{iopart}

\usepackage{graphicx}
\usepackage{xspace}
\usepackage{tabularx}
\usepackage{color}
\usepackage{amsbsy} 
\usepackage{xspace}

\newcommand{\bstss}{BiSbTeSe$_2$\xspace}

\newcommand{\bstsb}{Bi$_{1.5}$Sb$_{0.5}$Te$_{1.7}$Se$_{1.3}$\xspace}

\begin{document}

\title[]{Conduction spectroscopy of a proximity induced superconducting topological insulator}

\author{MP Stehno$^1$, NW Hendrickx$^1$, M Snelder$^1$, T Scholten$^1$ YK Huang$^2$, MS Golden$^2$, and A Brinkman$^1$}

\address{$^1$ Faculty of Science and Technology and MESA+ Institute for Nanotechnology, University of Twente, The Netherlands}
\address{$^2$ Van der Waals - Zeeman Institute, Institute of Physics, University of Amsterdam, The Netherlands}
\ead{a.brinkman@utwente.nl}
\begin{abstract}
The combination of superconductivity and the helical spin-momentum locking at the surface state of a topological insulator (TI) has been predicted to give rise to p-wave superconductivity and Majorana bound states. The superconductivity can be induced by the proximity effect of a an s-wave superconductor (S) into the TI. To probe the superconducting correlations inside the TI, dI/dV spectroscopy has been performed across such S-TI interfaces. Both the alloyed \bstsb and the stoichiometric \bstss have been used as three dimensional TI. In the case of \bstsb, the presence of disorder induced electron-electron interactions can give rise to an additional zero-bias resistance peak. For the stoichiometric \bstss with less disorder, tunnel barriers were employed in order to enhance the signal from the interface. The general observations in the spectra of a large variety of samples are conductance dips at the induced gap voltage, combined with an increased sub-gap conductance, consistent with p-wave predictions. The induced gap voltage is typically smaller than the gap of the Nb superconducting electrode, especially in the presence of an intentional tunnel barrier. Additional uncovered spectroscopic features are oscillations that are linearly spaced in energy, as well as a possible second order parameter component.
\end{abstract}

\maketitle

\section{Introduction}

The proximity effect between an $s$-wave superconductor and a three-dimensional topological insulator (TI) or nanowire with strong spin-orbit coupling has been predicted to create a stable zero-energy mode that may serve as the building block of a quantum computer, the Majorana zero-energy mode \cite{Kitaev01,Sau2010,Alicea2010,Ivanov01,Sarma2005,Nayak2008,Akhmerov09,qi11,Linder10,Tanaka09,Fu2009,Fu22009,Fu2008}. The presence of peaks in the conductance at zero-energy has been interpreted as experimental evidence for the existence of Majorana modes in such systems \cite{Deng2012,Mourik2012,Das2012,NadjPerge2014}. 

In order to create a Majorana zero-energy mode, theory requires a dominant (induced) $p$-wave component in the superconducting order parameter of the material. It is also known from theoretical studies that a zero-bias conductance peak and conductance dips at the characteristic gap energy are typical features of (dominant) $p$-wave superconductivity \cite{Sawa2007,Fu2008,Eschrig,Linder2010,Fu2010,Yamakage2012,Burset2014,Asano2013,Snelder2015}. Effective $s$-wave pairing with subdominant $p$-wave admixture has been predicted  \cite{Snelder2015,Tkachov2013,Tkachov2013b} at the interface between a topological insulator and an $s$-wave superconductor. However, as was shown by Tkachov in Ref.~\cite{Tkachov2013}, $s$ and $p$-wave correlations have a different spatial dependence suggesting that device geometry may play a crucial role in observing $p$-wave features in topological insulator/superconducting devices. It is, therefore, desired to investigate conductance spectra in proximized topological materials.

To study the proximity effect between an $s$-wave superconductor~({Nb}) and a 3D TI with dominant surface transport, we performed differential conductance measurements on Nb/TI/Au devices, where the TI is either alloyed \bstsb or stoichiometric \bstss. This generation of topological insulators has been shown before to have negligible bulk conductance in thin flakes \cite{Taskin2011,Xu2014}. The large difference in topological surface state mobility between the two types of TI provides different spectroscopic features such as a resistance peak due to electron-electron interactions in the disordered case and conductance oscillations in the cleaner, stoichiometric case. However, we reveal that the underlying spectra are very much alike. The general observations in the spectra of a large variety of samples are conductance dips at the induced gap voltage, combined with an increased sub-gap conductance, consistent with p-wave predictions. The induced gap voltage is typically smaller than the gap of the Nb superconducting electrode, especially in the presence of an intentional tunnel barrier. 

\section{Proximity effect into disordered \bstsb}
\subsection{Device fabrication}
All devices were prepared by exfoliation from a single crystal of \bstsb, grown by the Bridgeman method, which was found to exhibit vanishing bulk conductance in earlier experiments \cite{Snelder2014,Pan2015}. The flakes were transferred from the \bstsb crystal to a Si/SiO$_{2}$ substrate by means of mechanical exfoliation. By means of e-beam lithography (EBL) the Au electrode is defined on the flake. Thereafter we perform a low voltage 15 second Ar etching step to avoid large damaging of the surface followed by sputtering in-situ a 3 nm Ti layer and a 60 nm Au layer. The Ti layer is grown for better adhesion of the Au on the flake. Next, a second EBL step is performed to define the Nb part of the device. After the EBL step the structure is etched for 15 seconds again at low voltage followed by in-situ sputtering of 80 nm Nb. BSTS has a small mean free path of 10--40 nm. Therefore we optimized the EBL step so that a spacing of 50 nm between the two electrodes could be realized, in order to minimize the resistance contribution of the \bstsb interlayer itself. The width of the electrodes is about 300 nm, see also Fig. 1. The thickness of the exfoliated flakes is about 100 nm, which is thick enough to rule out a direct coupling between the top and bottom topological surfaces. Because of the very small electrode spacing at the top surface, we expect the contribution of the bottom surface to the spectra to be small. 

\subsection{Conductance spectroscopy}
The conductance was measured by applying a DC current with an additional AC current by means of a lock-in amplifier. We measured the differential voltage and the DC voltage across the devices. 

The conductance spectra at 1.7 K of two different devices, but with the same dimensions and geometry (samples D1 and D2) are shown in Fig. 2a. Although the line shapes of the devices look slightly different, two common features are readily identified. First of all, both spectra display dips in the conductance at a voltage that we will show is most likely related to the proximity induced gap value. The corresponding gap values are significantly lower in energy than the gap of the Nb electrode ($\Delta$ is about 1.3 meV for our Nb with a $T_c=8.4$ K). Secondly, the subgap conductance is increased with respect to the normal state resistance. 

The characteristic energy scale associated with the induced superconductivity is related to the dwell time in the junction and is known as the Thouless energy. In the diffusive limit it is defined as $\hbar D/L^{2}$ where $D$ is the diffusion constant \cite{Volkov1993,Belzig1996,Melsen1996}. The Thouless energy corresponds very well to energy scale attributed to the width of the dip features in all three the devices. This suggest that mainly diffusive transport contributes to the measured conductance spectra and that the observed features can be interpreted as signatures of induced superconductivity in \bstsb. 

\subsection{P-wave symmetry?}
One may raise the question whether or not the conductance dips at the gap value and the increased conductance near zero energy can be explained by the existence of $p$-wave superconducting correlations in the TI. Note, that a conductance dip near the (induced) gap energy is one of the characteristic signatures of $p$-wave superconductivity. At first sight, we do not expect a dominant $p$-wave order parameter to exist, as is noted in Ref. \cite{Snelder2015,Tkachov2013,Tkachov2013b}. Rather, an equal admixture of $p$ and $s$-wave correlations coexist. Nevertheless, as is shown by Tkachov \textit{et al.} \cite{Tkachov2013,Tkachov2013b} the amplitude of the $s$ and $p$-wave order parameters have different spatial dependencies. For a dirty TI (i.e. the coherence length $\xi_{N}$ is much larger than the mean free path $l_{mean}$), it is shown that the $s$-wave order parameter is always dominant over the $p$-wave and both decay to zero on a length scale of the order of the coherence length. However, for the clean limit or intermediate ratios of $\xi_{N}$ and $l_{mean}$, the $p$-wave order parameter can become dominant over the $s$-wave correlations, depending on the distance from the interface. For an estimated mean free path of 10 nm and a $T_{c}$ of 8.4 K we obtain $\xi_{N}=\sqrt{\frac{\hbar D}{2\pi K_{B} T_{c}}}=15.5$ nm ,which corresponds with $\xi/l_{mean}= 1.5$, or an intermediate regime according to Ref. \cite{Tkachov2013}. Now it depends on the distance over which the induced superconductivity is probed whether or not a dominant $p$-wave order parameter can be observed.

The existence of a zero-bias conductance increase and conductance dips at gap edges, as observed in our devices, are also known to appear in $s$-wave diffusive SNN' devices as described by Volkov \textit{et al.} \cite{Volkov1993}. It is shown that the total diffusion constant, directly related to the measured differential conductance, is given by
\begin{eqnarray}
D(E)=\frac{1+r_{1}+r_{2}}{r_{1}/M_{1}(E)+r_{2}/M_{2}(E)+m(E)},
\end{eqnarray} where $r_{1}$ and $r_{2}$ are the barrier resistances at the Nb/\bstsb and Au/\bstsb interface, respectively, normalized to the resistance of \bstsb, $M_{1}(E)$ is the spectral density in \bstsb at the position of the \bstsb/Nb interface, $M_{2}(E)$ the spectral density of \bstsb at the \bstsb/Au interface and $m(E)$ is the spectral density of \bstsb integrated over the distance between the two interfaces. If one of the interfaces is dominating it follows from this diffusion constant that the measured conductance is similar to the properties of this interface through which we can make an estimation of the distance over which the superconducting relations are probed. Device D2 has the largest modulation in the conductance and corresponds to 12\% of the total measured conductance. Following Volkov \textit{et al.} \cite{Volkov1993} this would correspond to a $r_{1}+r_{2}$ of the order of one. This implies that the different resistances are comparable to each other, which results in a measured conductance that is a weighted average of the three spectral densities and hence there is no dominant interface. This makes the conductance spectra rather complicated and prevents us from extracting a well-defined position from which one can conclude \cite{Tkachov2013,Tkachov2013b} whether $p$-wave correlations are dominant over the $s$-wave correlations.

\subsection{Disorder induced zero-bias resistance peak}

In order to distinguish between the two scenarios for the conductance spectra, spectroscopic features can be investigated as function of temperature. For this purpose, another device (D3) was cooled down further, see Fig. 2b. The position of the induced gap values clearly changes as function of temperature, as one would expect. However, the zero-bias conductance increase (dashed line around zero voltage in Fig. 2b) now becomes masked by a zero-bias resistance peak at temperatures below 1.5 K. We attribute the zero-bias resistance peak to disorder induced electron-electron interactions (EEI, also called the Altshuler-Aronov effect) since this feature is also there in the absence of superconducting electrodes (e.g. see our reference junctions in Ref. \cite{Tikhonov2016}). Moreover, the interplay between EEI and the related weak antilocalization (WAL) is a well studied phenomenon in topological insulators \cite{Wang2011,Lu2014}. In our case, the voltage here determines the relevant energy scale instead of temperature. The logarithmic dependence is expected for two-dimensional EEI \cite{Wang2011}, whereas a square root dependence would have been expected for a three-dimensional situation. The zero-bias conductance peak in a perpendicular magnetic field (see Fig. 2c) in related Nb-\bstsb-Nb junctions (samples S1 and S2) with the same geometry, can be best fitted (cross section at a field of 1 T in Fig. 2d) with a logarithmic dependence on voltage. 

In order to further explore the p-wave symmetry as an explanation for the spectra, one has to get rid of the strong disorder that causes the resistance peak at low temperatures. Therefore, we now turn to stoichiometric \bstss, with higher surface state mobility. Furthermore, to enhance the spectroscopic signal from the interface with respect to the topological layer, we employ edge-type side contacts with intentional tunnel barriers.

\section{Proximity effect into stoichiometric \bstss}

\subsection{Device fabrication}
To make edge-type side contacts, flakes were again transferred from the stoichiometric \bstss crystal to a Si/SiO$_2$ substrate by mechanical exfoliation. The big drain contact is defined by optical lithography. Next, a short, low power Ar etch is performed, to enhance the transparency of the contact, followed by the sputter deposition of 5 nm of Pd, a 100 nm Nb layer, and a 5 nm Pd capping layer. Thereafter, the edge is defined by optical lithography and the flake is Argon ion milled down. A short wet etch in nitric acid is performed to remove part of the disordered material, formed by the ion milling. The (crosslinked) photoresist defining the edge is not removed, so the top of the flake remains covered. A third optical lithography step is used to define the edge contacts. The exact alignment is not important, since the crosslinked photoresist protects the top of the flake, as is visible in Fig. 3a. The edge contacts are sputter deposited under an angle of 45$^{\circ}$, to ensure good side wall coverage. First, a layer of aluminium is deposited and oxidised, to form a tunnel barrier, followed by a 100 nm Nb layer and a 5 nm Pd capping layer. Finally, both layers of photoresist are dissolved, causing the metal layer to break at the top of the flake, leaving a side contact, as shown in Fig. 3b and the image shown in Fig. 3c. The exfoliated flake thicknesses range between 50 and 300 nm. Since superconductivity can also be induced from the side into the bottom topological surface, the conductivity spectra could have parallel contributions from the top and bottom surfaces.

\subsection{Conductance spectra at renormalized voltage}

Both the bias voltage as well as the differential resistance are measured between the current sourcing contact and the big drain contact. Therefore, this voltage drops not only over the side contact interface, but also over the flake itself as well as the big drain contact interface. Since the flake is only a thin sheet of material and BSTS is not doped, this resistance can actually have a significant value. In order to estimate the flake resistance, it is possible to solve Poisson's equation for a piece of material shaped like this flake.

By modeling the Poisson equation for the geometry of the sample (see Fig. 3c) and by assuming that the drain contact is transparent (because of the relatively large size and absence of barrier), the potential profile was numerically calculated. An example is given in Fig. 3d, for conduction across the top contact. Using the TI resistivity and by combining the measurement results of local and non-local configurations (i.e. measuring the potential at other contacts then the current injectors), an estimate for the ratio between interface resistance and interlayer resistance was made. Using the estimated flake resistances, it is possible to correct the bias voltage for the voltage drop across the flake, simply by considering the flake and contact to form a voltage divider. The correction factors are found to be $f_{C1}=0.431$ and $f_{C2}=0.589$. Now the $dI/dV$ curves can be depicted as function of the estimated bias voltage across the side contact interface only, as can be seen in Fig. 4a. Note that using these correction factors, the sizes of the small gaps of both devices C1 and C2 match, being approximately $0.2$ mV, as well as many of the other peaks as will be discussed later on. 

The conductance spectra of samples C1 and C2 resemble the spectra of the more disordered samples D1 and D2. However, the Nb/\bstss interface now has a stronger weight in the spectra and the mean free path in \bstss is higher compared to the more disordered \bstsb. It is, therefore, more plausible now that the spectra can indeed be explained by dominant p-wave correlations \cite{Tkachov2013}. Now, that less disorder is present, we do not expect EEI to become important, and we can study the induced gap values as function of temperature. The results are plotted in Fig. 4b. Upon cooling down, it can be seen that the induced gap opens up around 1.3 K, and then quickly increases in size, reaching its final width at 0.8 K. Note, that upon cooling down even further, another small conductance peak at zero bias seems to arise. The size of the induced gap is plotted versus the temperature in the inset, together with the expected Bardeen-Cooper-Schrieffer gap size dependence of a superconductor with a critical temperature of 1.4 K. There is a good agreement between the two. 

\subsection{Interference effect}
The differential resistance spectra show clear oscillations with a period that is constant in bias voltage. Assigning a peak number to the dips in the spectra of sample C1, starting from zero bias, the renormalized bias voltage positions of the oscillation peaks are plotted versus the peak number in Fig. 4c. The same was done for sample C2, but the assumption was made that occasionally a peak is missing for C2, in order to match peaks/dips of different contacts at the same bias voltage. The similarities between the two contacts are striking. The resonances are present throughout the entire range of bias voltages, both within the induced gap, as well as out of the gap. The transition between two regions with linearly spaced peaks is around 0.5 mV, somewhat outside the induced gap value. The difference in the period between these two domains is about a factor of 4. These oscillations cannot be detected on any of the nonlocal contacts, implying that they occur in or near the local, current sourcing, junction.

There can be many causes of oscillations in tunnel junctions. Multiple Andreev reflections (MAR) would cause a subgap interference pattern with maxima at $eV=2\Delta/n$ \cite{Klapwijk1982}. The oscillations we observed are linearly spaced, excluding MAR as a possible reason. We also rule out de Gennes-Saint James resonances \cite{DeGennes1963}. These would be almost linearly spaced in voltage, but they require the absence of states within the gap (good interface), as indicated in Refs. \cite{Koren1996, Deutscher2005}. As the zero bias background conductance in our device does not vanish inside the gap, these kind of oscillations likely cannot be present. Also, self-induced steps from internal coupling to Josephson radiation emission would require a Josephson junction to be formed across the tunnel barrier, which we do not resolve experimentally. 

Tomasch oscillations are geometrical resonances on the superconductor side of the interface \cite{Tomasch1966}. When an electron-like (hole-like) quasiparticle Andreev reflects into a hole-like (electron-like) quasiparticle on the interface, it can immediately interfere with itself, due to the mixed nature of the quasiparticle state. This results in a series of oscillations with peaks in the conductance at $eV_p=\sqrt{\Delta^2+\left(\frac{nhv_F'}{2d_S}\right)^2}$, with $n$ the oscillation index, $v_F'$ the renormalised Fermi velocity in the superconductor and $d_S$ its thickness. This will thus result in an almost linear series of resonances in the differential conductance spectrum. Assuming a Fermi velocity of $6\cdot 10^5$ ms$^{-1}$ \cite{Taskin2011} in the \bstss, the outer set of resonances can only be fitted assuming a superconductor thickness of about $5\ \mu$m. This does not match any dimensions present in our device.  We thus conclude that Tomasch oscillations do not seem to be the origin of the observed resonances. Similarly we rule out McMillan-Rowell oscillations \cite{Rowell1966} inside the flake (i.e. between the side contact and the drain), since this would require phase coherence and ballistic transport across at least two times the width of the flake, i.e. $>10\ \mu$m, so this is highly unlikely. Note, that the resonances in S-TI-S junctions by Finck \textit{et al.} \cite{Finck2014, Finck2016} are in fact attributed to Fabry-Perot resonances in the TI, but there the interlayer thickness is a lot shorter. Furthermore, if the resonances were to occur in our flake, they should also be present in the nonlocal signal. This is not the case in our measurements.

Since the interference does not seem to match any geometric resonance, the origin of the oscillations could perhaps be sought in interference between multiple order parameter contributions. Self-consistent calculations of the order parameter of a TI on top of a superconductor have been performed by Dellabetta \textit{et al.} \cite{Dellabetta2014}, and show oscillations in the order parameter in $k$-space . If the size of the order parameter was to fluctuate, the magnitude of the gap (and thus the DOS) would also show oscillations for different $k$-vectors. It could be that the oscillations we observed, are caused by similar effects, related to multiple order parameters interfering.

Speculatively, another indication for the presence of two order parameters is provided by the temperature dependence of the spectra at higher temperatures. A different device (C3) was measured above 1.7 K, see Fig. 4d. Note, that in this case, the bias voltage shown is not corrected for the voltage drop over the \bstss. By comparing the shape of the curve at 1 K with the 1 K data of samples C1 and C2, the zero bias conductance increase can readily be associated with the induced gap (rescaled value) that is plotted as function of temperature in the inset of Fig. 4a. However, at higher temperatures, the spectra of sample C3 show a dominant decrease of the conductance around zero bias, with an associated gap value that is about twice as high in energy. This could perhaps indicate the presence of an additional order parameter component.

\section{Discussion}

The observed conductance spectra are consistent with an unconventional order parameter symmetry, such as the predicted p-wave contribution. Also Yang \textit{et al.} have reported the observation of a conductance increase at low energy in a Sn-Bi$_2$Se$_3$ junction \cite{Yang2012}, which they attributed to an induced unconventional order parameter. Furthermore, they have also observed resistance spikes at nonzero energy, similar to our observations. Koren \textit{et al.} also observed a zero-bias conductance peak in a proximity induced superconducting phase in a TI \cite{Koren2011}. They find a conductance increase in conductance spectroscopy measurements of a pure bismuth film, which they attribute to induced superconductivity, due to islands of bismuth becoming superconducting. Argon milling is known to disorder the surface of a material extensively, so a similar effect could, in principle, also occur in our junctions. In Ref. \cite{Koren2015} Koren et al. show that proximity superconductivity can really be realised in Bi$_2$Se$_3$, by observing an increase of the conductance of grainy NbN-films when a Bi$_2$Se$_3$-film is deposited on top. Many publications on proximity induced superconductivity into Bi$_2$Se$_3$ report the proximity induced superconductivity to appear at a temperature below the superconducting electrode $T_c$, similar to our findings \cite{Koren2011,Koren2013,Qu2012}. Recently, also the proximity effect into topological HgTe has been studied by conductance spectroscopy. 

In summary, all spectroscopic devices studied here show spectra that are reminiscent of p-wave superconductors, i.e. having conductance dips at the value of the induced gap, as well as an increase in conductance below the gap. The stoichiometric \bstss is more suited for spectroscopy than the disordered \bstsb, in which case electron-electron interaction induced zero-bias resistance peaks can mask the actual conductance increase. The presence of oscillations as well as the possible observation of a second order parameter component are experimental observations that can, hopefully, provide input to theoretical modeling of the induced proximity effect.

\newpage

\begin{figure}[t!]
	\centering 
		\includegraphics[width=1\textwidth]{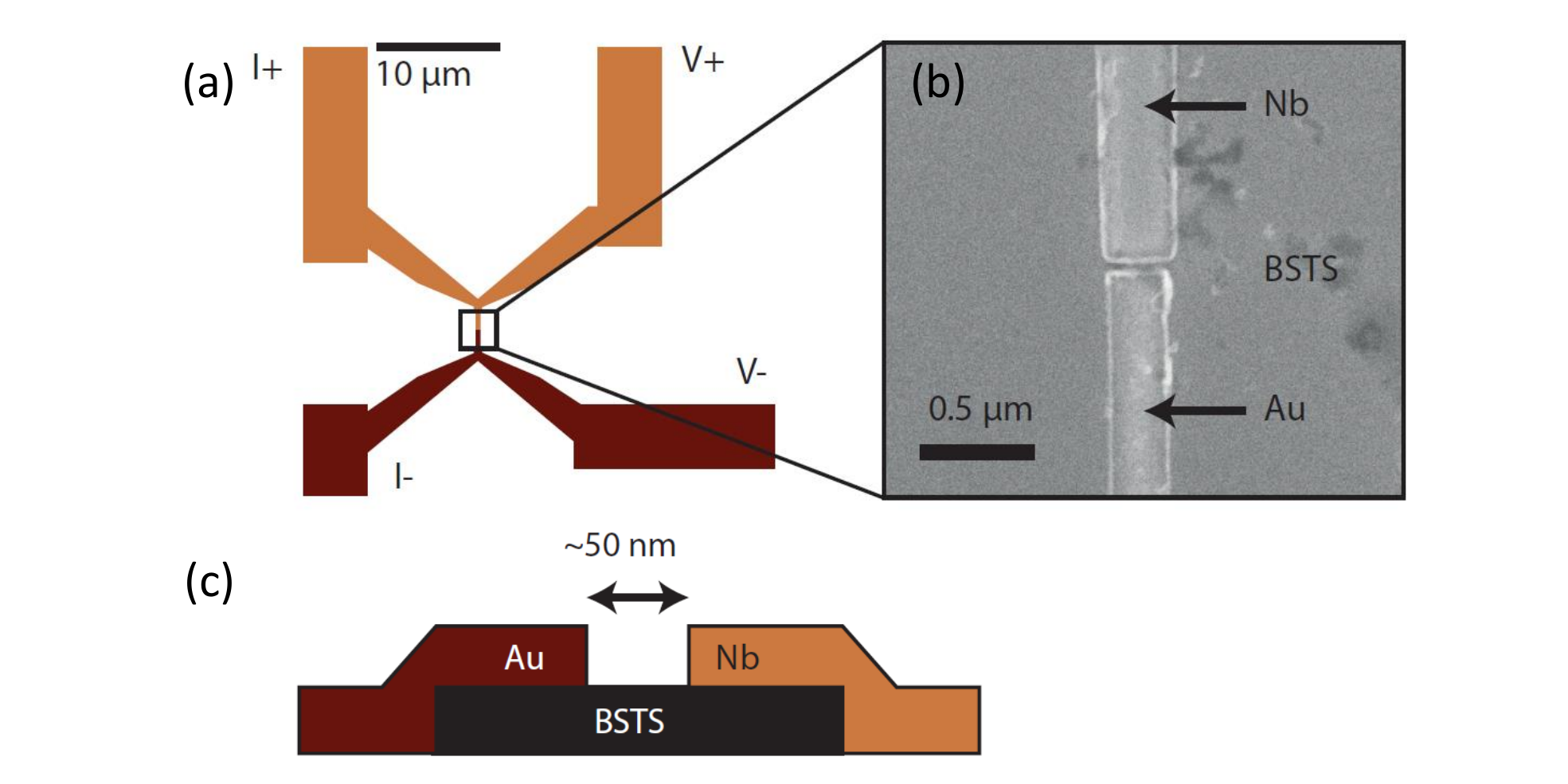}
		\caption{(a) Schematic view of the Nb-\bstsb-Au spectroscopic devices, indicating the current and voltage contacts. (b) Scanning electron microscopy image of the device. The spacing between the Au and Nb electrodes is about 50 nm and the width of the contacts is about 300 nm. (c) Schematic side view. In this geometry, the Nb induces superconductivity in the topological insulator underneath. The proximity effect extends laterally beyond the Nb electrode.} 
		\label{fig1}
\end{figure}

\begin{figure}[t!]
	\centering 
		\includegraphics[width=1\textwidth]{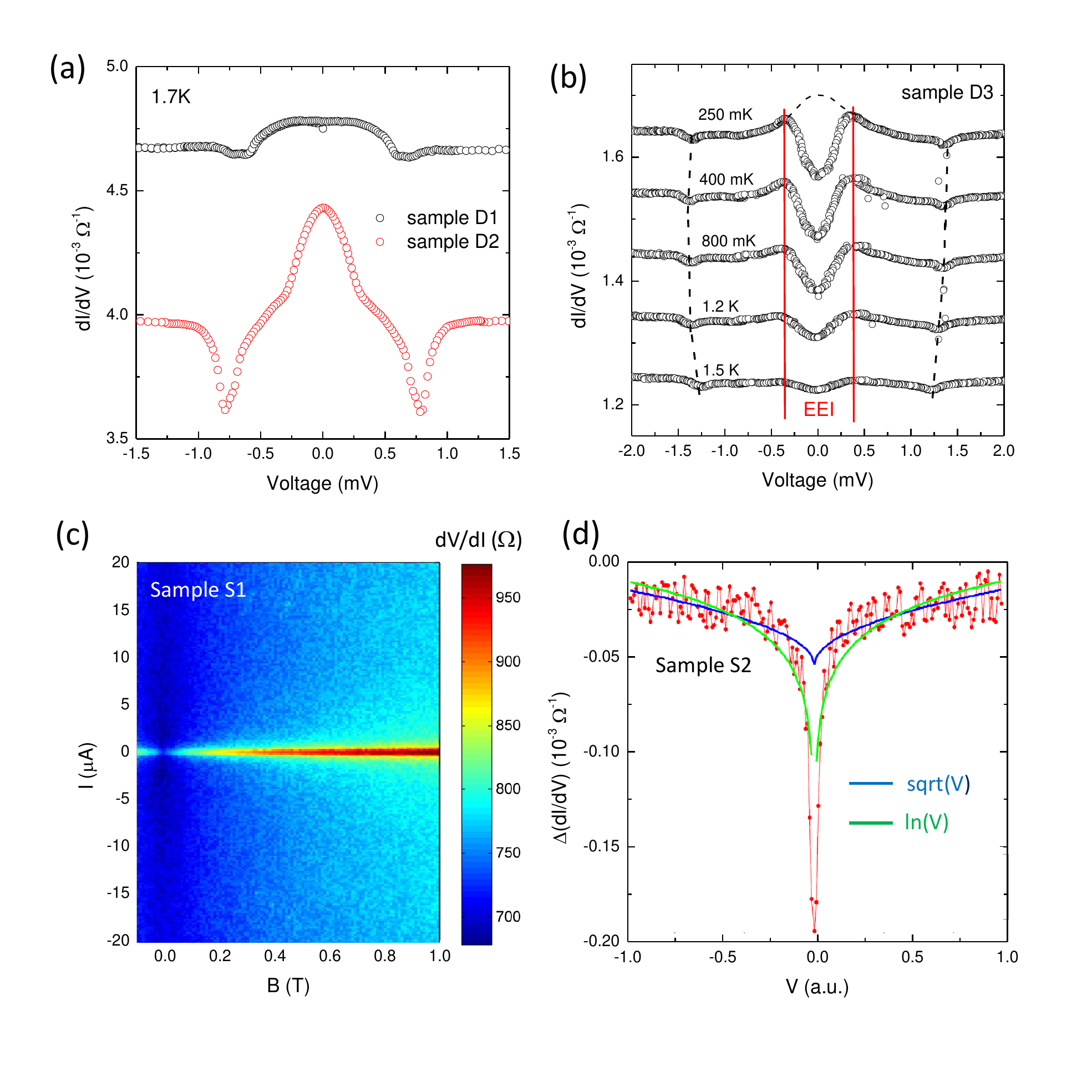}
		\caption{(a) Conductance spectra of two different Nb-\bstsb-Au spectroscopic devices, with disordered interlayer, samples D1 and D2, at $T=1.7$ K. Generic features are conductance dips at some characteristic gap voltage and an increased conductance around zero bias. (b) Conductance of Nb-\bstsb-Au device D3 at various temperatures. For clarity, the data have been off-set in steps of $0.1 \times 10^{-3} \Omega^{-1}$ with respect to the data at 1.5 K. The black dashed lines are guides to the eye to see how the gap features move to lower voltage for increased temperature. The red lines indicate the onset of a disorder induced electron-electron interaction (EEI)resistance peak, masking the conductance increase at zero bias as observed at higher temperature for samples D1 and D2. (c) Sample S1 is a Nb-\bstsb-Nb device with electrode spacing of 250 nm. The EEI resistance peak at zero-bias is plotted as function of bias current and the magnetic field applied perpendicular to the sample. (d) The conductance at a field of 1 T of a similar Nb-\bstsb-Nb device with electrode spacing of 200 nm shows a logarithmic dependence on voltage (best fit: green line), as expected for EEI on a two-dimensional surface. Three-dimensional EEI would have given a square root dependence (best fit: blue line).  }
		\label{fig2}
\end{figure}

\begin{figure}[t!]
	\centering 
		\includegraphics[width=1\textwidth]{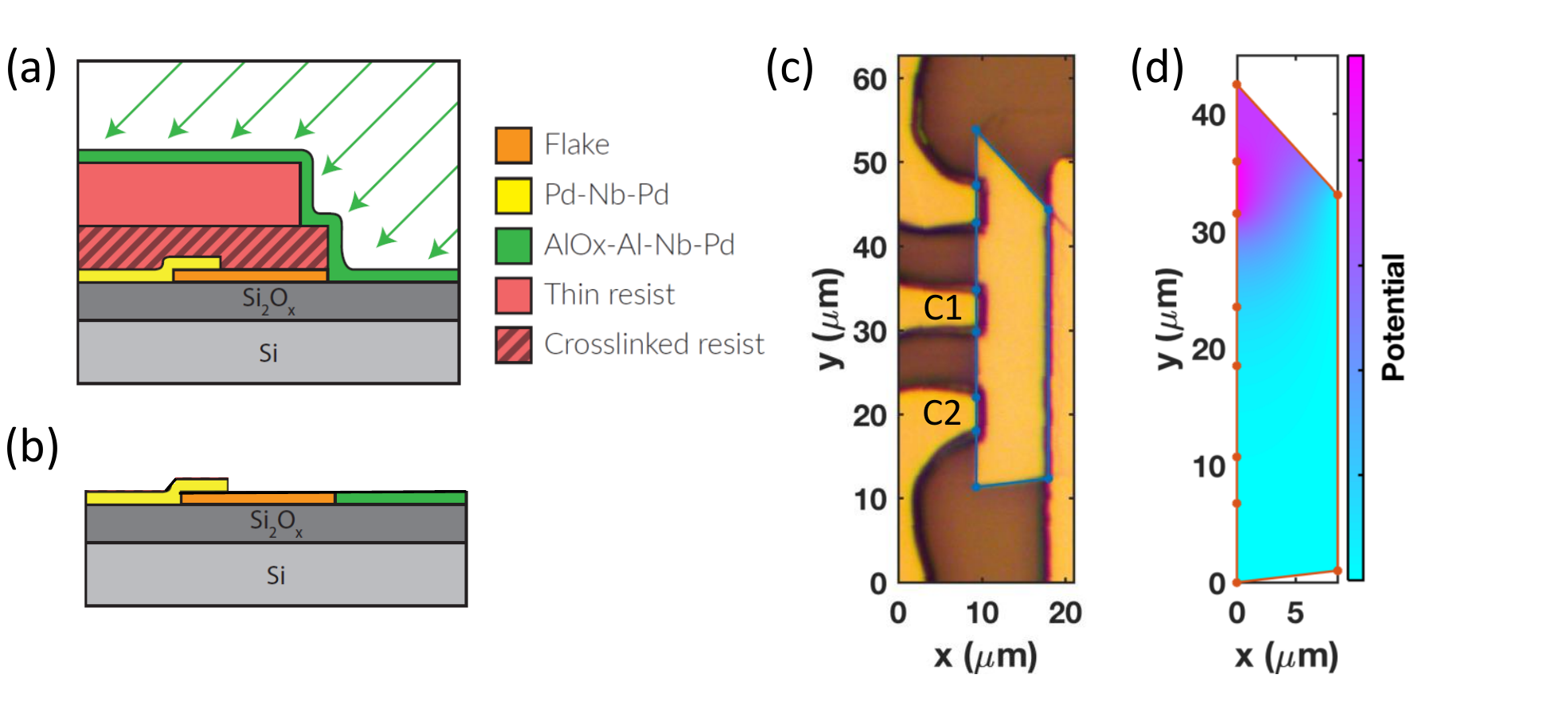}
		\caption{(a) Schematic drawing of one of the fabrication steps of Nb/Al$_2$O$_3$-\bstss-Pd/Nb/Pd devices with stoichiometric interlayers. The Nb electrode deposition is preceded by a thin layer of oxidized Al to form a tunnel barrier, effectively increasing the role of the Nb-\bstss interface in the conductance spectra. (b) Lay-out after the final fabrication step. The Nb-\bstss contact is in the plane of the topological surface state, rather than perpendicular to that. (c) Optical microscopy image of a sample with the \bstss devices C1 and C2. The geometry used for calculating the potential distribution is overlaid on the image. (d) Calculation result of the electric potential model when the device is biased between the right electrode and the top left electrode. Similarly, for the contacts C1 and C2, the contribution to the resistance of the interlayer was estimated.} 
		\label{fig3}
\end{figure}

\begin{figure}[t!]
	\centering 
		\includegraphics[width=1\textwidth]{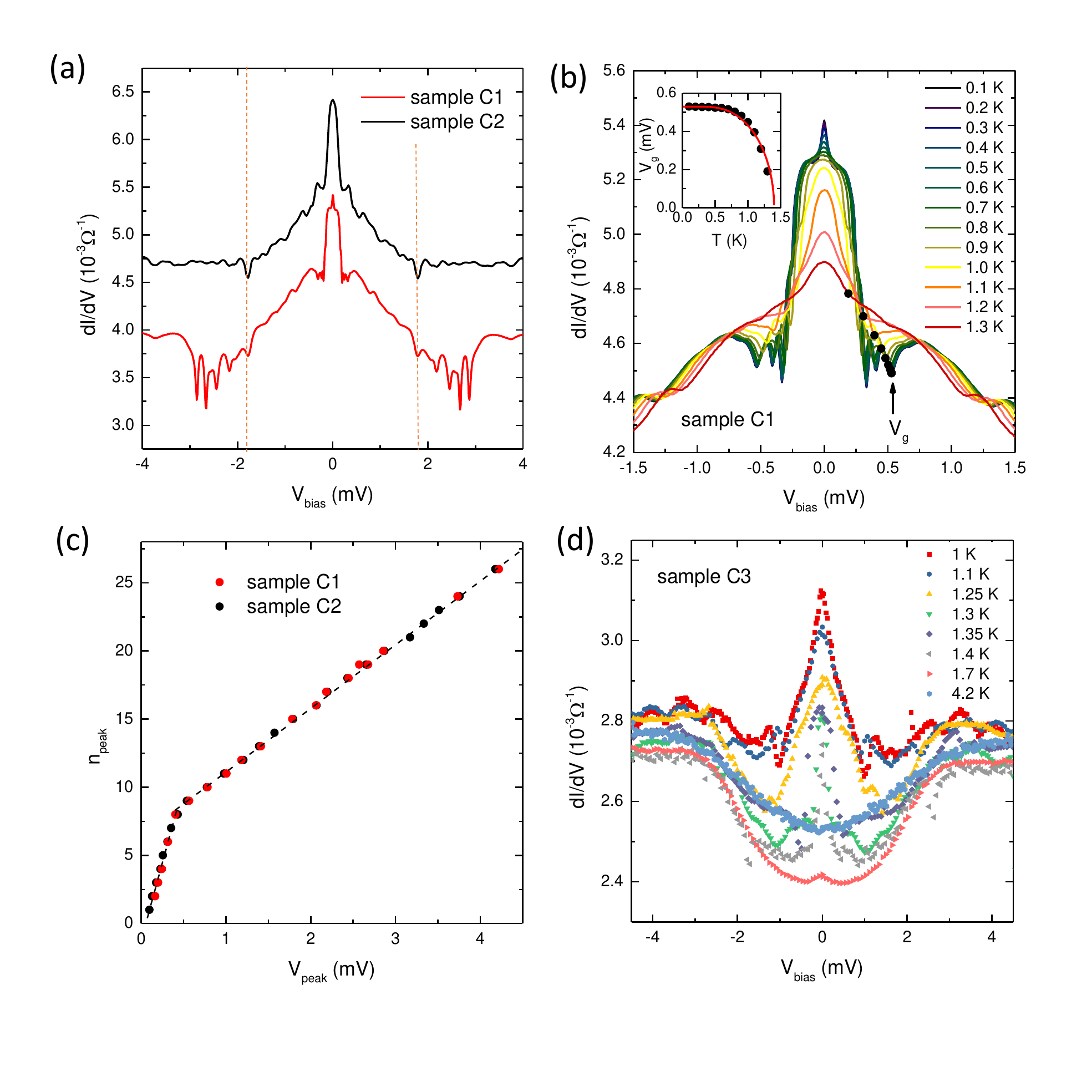}
		\caption{(a) Conductance spectra of \bstss sample C1 and C2 at 15 mK. After correcting for the voltage across the interlayer, the spectroscopic features (gap as well as interference peaks) of the same sample seem to coincide quite well. (b) Conductance spectra of \bstss device C1 at various temperatures. The bias voltage has been corrected for the potential across the interlayer. Conductance dips at a characteristic gap voltage as well as a conductance enhancement at zero bias are present in these devices too. The position of the gap feature (black circles) has been plotted as function of temperature in the inset. The red solid line shows a fit to the Bardeen-Cooper-Schrieffer gap equation for a superconductor with a critical temperature of 1.4 K. (c) The voltage at which peaks/dips are observed in the conductance spectra of samples C1 and C2 at 15 mK are plotted as function of the peak position. (d) Sample C3 (\bstss device on a different crystal flake) is measured at temperatures up to 4.2 K. The voltage axis has not been corrected for the contribution from the interlayer. Besides the conductance dips and zero-bias conductance enhancement, at higher temperature, a dominant conductance decrease is revealed around zero bias, with a cut-off that is higher in energy.  } 
		\label{fig4}
\end{figure}

\end{document}